\documentclass[a4paper,12pt]{article}
\usepackage{graphics}
\usepackage{amsfonts}
\usepackage{epsfig,amssymb,amsmath,graphicx,caption,subcaption,
verbatim,hyperref,xcolor,ulem,epstopdf,psfrag,pstool,braket,array,
enumerate,wrapfig,tabularx,setspace,geometry}

\begin{document}
%

\newcommand{\C}{\mathbb{C}}
\newcommand{\CP}{\mathbb{CP}}
\newcommand{\bP}{\mathbb{P}}
\newcommand{\RP}{\mathbb{RP}}
\newcommand{\spp}{\mathbb{S}}
\newcommand{\R}{\mathbb{R}}
\newcommand{\T}{\mathbb{T}}
\newcommand{\Z}{\mathbb{Z}}
\newcommand{\OO}{\mathcal O}
\newcommand{\PP}{\mathbb{P}}
\newcommand{\Tr}{\mbox{Tr}}
\newcommand{\edth}{{\p \!\!}'}
\newcommand{\koniec}{\begin{flushright}  $\Box $ \end{flushright}}

\newcounter{mnotecount}[section]
\renewcommand{\themnotecount}{\thesection.\arabic{mnotecount}}

\newcommand{\mnotex}[1]
{\protect{\stepcounter{mnotecount}}$^{\mbox{\footnotesize
$
\bullet$\themnotecount}}$ \marginpar{
\raggedright\tiny\em
$\!\!\!\!\!\!\,\bullet$\themnotecount: #1} }

\newcommand{\mnote}[1]
{\protect{\stepcounter{mnotecount}}$^{\mbox{\footnotesize
$
\bullet$\themnotecount}}$ \marginpar{
\raggedright\tiny\em
$\!\!\!\!\!\!\,\bullet$\themnotecount: #1} }
\newcommand{\md}[1]{\mnote{{\bf MD:}#1}}

\def\be{\begin{equation}}
\def\ee{\end{equation}}
\def\p{\partial}
\def\ov{\overline}
\def\ll{\lambda}
\def\theequation{\thesection.\arabic{equation}}
\thispagestyle{empty}
\vskip 3em
\begin{center}
{{\bf  \Large Equivalence principle, de-Sitter space, and cosmological twistors}} 
\\[15mm]

{\bf \large Maciej Dunajski\footnote{email: M.Dunajski@damtp.cam.ac.uk}}\\[20pt]

\vskip 1em
{\it 
Department of Applied Mathematics and Theoretical Physics,\\
University of Cambridge, \\
Wilberforce Road, Cambridge CB3 0WA, U.K.}\\
\vskip 1em
\date{March 2023}

\abstract
{   I discuss  the impact of the positive cosmological constant on the 
interplay between the equivalence principle in general relativity, and the rules of quantum mechanics.  
At the non--relativistic level there is an ambiguity in the definition of a phase of a wave function
measured by inertial and accelerating observes. This is the cosmological analogue of the Penrose effect, which
can also be seen as a non--relativistic limit of the Unruh effect.  The symmetries of the associated Schr\"odinger equation are generated by the Newton--Hooke algebra, which arises from a non--relativistic limit of a cosmological twistor space. }
\end{center}
\vskip 30pt
First award winning essay of the  2023 Gravity Research Foundation
competition.

\vfill
\newpage
\setcounter{page}{1}
\renewcommand{\thefootnote}{\arabic{footnote}}

\section{Introduction}
According to the historical account \cite{history}, along the inverse square law of gravitation Newton also considered gravitational forces between two bodies which vary linearly with the distance. A superposition 
of these two  forces results from a radial potential
\be
\label{NG}
V=A/r+B r^2.
\ee
This, with $B=0$ and $A=-GM$ where $G$ is the gravitational constant, is the gravitational potential outside a sphere of mass $M$. If $A=0$ and $B$ is positive then $V$ gives the Hooke law of elasticity.
We will be referring to $V$ of the form (\ref{NG}) as the Newton--Hooke potential even if $B$ is allowed to be negative.

The current observational evidence is that  the Universe is in a stage of accelerated cosmic expansion in agreement
with the presence of a very small positive cosmological constant $\Lambda$. The Newtonian potential resulting
from a combination of this cosmological term with the inverse square law is
of the form (\ref{NG}) with $B=-\frac{1}{2}\omega^2$. This 
arises by taking a non--relativistic limit of the Schwarzchild--de Sitter metric, where the speed of light $c\rightarrow\infty$ and the cosmological constant $\Lambda\rightarrow 0$, but 
$
\omega^2\equiv\frac{1}{3}{c^2\Lambda}
$
remains finite.

The aim of this essay  is to discuss the impact of the cosmological term (however small it may be) on the 
interplay between the equivalence principle in general relativity, and the rules of quantum mechanics. If $\Lambda=0$, and the gravitational
field is uniform, then the Schr\"odinger equation can either be considered with a linear
 potential, or in the free falling 
frame with zero potential. The two frames are related by a coordinate 
transformation quadratic in time, and the 
corresponding wave functions differ by a phase factor with a cubic dependence on time. This 
non--linear time dependence in the phase is also present for  non--uniform  fields \cite{DP}, and 
poses a problem for the standard QFT framework,  as it leads to ambiguities in a definition of positive and negative frequency 
if a quantum superposition of two massive objects is considered \cite{Pe11}. 
In \S\ref{secu} we will demonstrate that this ambiguity arises as a non--relativistic limit of the Unruh effect \cite{unruh}.
With $\Lambda >0$, the analog of the uniform gravitational field 
leads to  the Schr\"odigner equation with a non--isotropic reversed harmonic oscillator potential. The  
phase ambiguity
has an
effect of shifting the cosmological horizon between  an inertial and an accelerating observer.

Our computational tool in \S\ref{qphase} is the Eisenhart approach
\cite{eisenhart, duval2}
which allows to deduce
the nonrelativistic mechanics in $3$ space and $1$ time dimensions with a potential (\ref{NG})  from the properties of a curved plane wave metric in $(4, 1)$
dimensions.
In this framework 
the classical trajectories are the projections of null geodesics, and the quantum wave functions correspond
to complex solutions of the wave equation which scale with a constant
factor under a null translation. The case $B=0$ with large $r$ leads to a flat $pp$--wave 
and explains the cubic--in--time phase ambiguity
\cite{horvathy, Silagadze, DP}.
If instead $A=0$  the potential corresponds to the non--relativistic limit of the de-Sitter space. The $pp$-wave
metric is then conformally flat, and a transformation to the Beltrami coordinates leads
a non--unitary transformation between a Schr\"odinger wave function of a free particle,
and that of a a particle moving in a reversed isotropic oscillator potential.
The general case corresponds, for large $r$,  to the $pp$-wave of constant curvature, and a non--isotropic reversed oscillator.

Finally  in \S{\ref{section2}} we discuss the cosmological twistor space of the de Sitter 
space, and  show  how the Newton--Hooke
algebra arises from a non--relativistic limit in terms of 
holomorphic vector fields in the twistor space.
\section{Schwarzchild--de-Sitter and its non--relativistic limit}
\label{section1}
The $(3+1)$ dimensional de--Sitter space is the maximally symmetric Lorentzian manifold
with positive scalar curvature which arises as a one--sheeted hyperboloid
in the $(4+1)$--dimensional Minkowski space.
There are several coordinate forms of  de-Sitter metric, and the one relevant for our 
considerations is
\be
\label{dS}
ds^2= c^2 f  dt^2-f^{-1}dr^2-r^2 h_{S^2},\quad\mbox{where}\quad f=1-\frac{\Lambda r^2}{3}
\ee
and $h_{S^2}$ is the round metric on a unit two--sphere.
The static coordinates $(r, t)$ only cover a part of de-Sitter space bounded by the cosmological horizon
$r=\sqrt{3/\Lambda}$. The regions outside this horizon can not be probed by a single observer. This cosmological horizon is observer-dependent, which will play a role below, when we argue that this leads to ambiguities in the quantum phase once the principle of equivalence is taken into account.

To formulate a non--relativistic quantum mechanics in the de-Sitter space one needs to take its non--relativistic limit.
We will do it for the more general case of the Schwarzchild-de-Sitter metric,
which is still of the form (\ref{dS}), but with 
\[
f=1-\frac{2M}{c^2r}-\frac{\Lambda r^2}{3},
\]
where $M$ a non--negative constant. The Schwarzchild metric has $\Lambda=0$, and the de-Sitter metric has $M=0$. The limit we are going to consider corresponds to simultaneously taking $c\rightarrow\infty$ and $\Lambda\rightarrow 0$, but such that the combination
\[
\omega^2\equiv \frac{\Lambda c^2}{3}
\]
is fixed, and remains finite. The current observational value of the cosmological constant gives $\omega^2\sim 10^{-35}s^{-2} $. The metric (\ref{dS}) blows up in the limit, but the Christoffel symbols stay finite, with the only non--zero components given by ${\Gamma^j}_{tt}=\delta^{jk}\p_k V$, where
\be
\label{NHP}
V=-\frac{GM}{r}-\frac{1}{2}\omega^2 r^2
\ee
is the Newton--Hooke potential (\ref{NG}), where $(A, B)$ have now been fixed.
\section{Quantum phase} 
\label{qphase}
The de-Sitter space does not admit a globally defined time--like Killing vector. Any generator of the de Sitter group $SO(1, 4)$ has to be space--like in some regions, for example close to the space--like conformal  infinities. This leads to problems with conventional formulations of quantum mechanics, as stationary states can not be globally defined. Additional problems result from the existence of cosmological horizon which is observer dependent. This, as argued by Gibbons and Hawking \cite{GHds}, rules out the existence of the $S$--matrix. We shall consider another set of difficulties which result from applying the equivalence principle
to a quantum particle moving in a uniform gravitational field.
 
 A convenient tool in making coordinate transformations between different frames of reference in the Schr\"odinger
equation is  the Eisenhart metric in $(4+1)$ dimensions 
\be
\label{eisenhart}
G=2 dudt +2 \frac{V({\bf x}, t)}{m} dt^2- d {\bf x}\cdot d{\bf x}
\ee
where $m$ is a constant with the dimension of mass. 
The Schr\"odinger equation then arises from the complex solution
$\phi$ of the wave equation on $G$:
\be
\label{wave_sh}
\mbox{If}\;\;\phi(u, t, {\bf x})=e^{-\frac{imu}{\hbar}}\psi(t, {\bf x}), \quad \mbox{then}\quad
 \Box_G{\phi}=0  \;\;\mbox{iff}\;\;   i\hbar \frac{\p \psi}{\p t}=-\frac{\hbar^2}{2m}\Delta \psi+V\psi.
\ee
\subsubsection*{The uniform gravitational field}
If $V=-m{\bf g}\cdot{\bf x}$ then the Eisenhart metric is flat, and in the flat coordinates
\be
\label{coords1}
T=t, \quad U=u-t{\bf g}\cdot {\bf x}+\frac{1}{6}|{\bf g}|^2t^3, \quad {\bf X}={\bf x}-\frac{1}{2}{\bf g}t^2
\ee
it takes the form
$
G=2dUdT-d{\bf X}^2.
$
Expressing a solution of the wave equations in both coordinate systems
$
\phi=e^{-\frac{imU}{\hbar}}\Psi({\bf X}, T)=e^{-\frac{imu}{\hbar}}\psi({\bf x}, t)
$
yields 
\be
\label{phase1}
  \psi({\bf x}, t)=\Theta({\bf x}, t)\Psi\Big({\bf x}-\frac{1}{2}{\bf g}t^2, t\Big), \quad\mbox{where}\quad
\Theta=e^{-\frac{im}{\hbar}\Big(\frac{t^3g^2}{6}-tgz\Big)}.
\ee
The function $\Psi({\bf X}, T)$ satisfies the free Schr\"odinger equation, and is  related to $\phi$
by a unitary transformation, but the phase $\Theta$ depends on $t$ in the non--linear way 
\cite{horvathy, Silagadze, DP}.
\subsubsection*{Pure cosmological term}
If $V=-\frac{1}{2}m\omega^2 r^2$, then the Eisenhart metric is conformally
flat, and takes the form \cite{GP}
\begin{eqnarray}
\label{cord_change}
G&=&\Omega^2(2dUdT-dR^2-R^2 h_{S^2}), \quad\mbox{where}\quad \Omega^2=\frac{1}{1-\omega^2 T^2},\quad\mbox{and}\nonumber\\
T&=&\frac{1}{\omega}\tanh{(\omega t)}, \quad R=\frac{1}{\cosh{(\omega t)}}r, \quad
U=u-\frac{1}{2}\omega r^2\tanh{(\omega t)}.
\end{eqnarray}
Let $\psi$ be a solution of the Schr\"odinger equation with the reversed isotropic harmonic oscillator potential 
$V$ in the $(r, t)$ coordinates. Therefore (\ref{wave_sh})
is in the kernel of the wave operator of $G$. The conformal invariance of the wave operator in dimension $5$ implies that
$
\Phi=\Omega^{3/2}\phi
$
satisfies the wave equation on the flat space time with the flat metric 
$\hat{G}=\Omega^{-2}G$. Writing
$\Phi=e^{-\frac{imU}{\hbar}}\Psi({\bf X}, T)$, and using (\ref{cord_change}) we find a
non-unitary transformation (a hyperbolic version of the Niederer transformation \cite{niederer})
between a wave function in a reversed harmonic oscillator potential, and that
of a free particle
\[
\psi(r, t)=(\cosh{\omega t})^{-\frac{3}{2}}
\Theta(r, t)\Psi(R, T),\quad\mbox{where}\quad
\Theta=e^{\frac{im}{\hbar}\Big(\frac{1}{2}\omega r^2\tanh{\omega t}\Big)}.
\]
\subsubsection*{Uniform gravitational field on cosmological background}
Consider the Newton--Hooke potential (\ref{NHP})
and expand it, up to quadratic terms, around the point
${\bf x}_0=(0, 0, R)$ in a coordinate
$z={\bf k}\cdot {\bf x}$, where ${\bf k}=(0, 0, 1)$ defines the direction of the
uniform field. Truncating the expansion at the quadratic order, 
completing the square and shifting $z$ to a new coordinate $Z$, yields $V=V_0-\frac{1}{2}{K^2} Z^2$, where $V_0$ is a constant, and $K^2$ is modified from $\omega^2$ by the presence of the mass $M$.
At the level of the Eisenhart metric (\ref{eisenhart})
the constant term in the potential can be eliminated by
a time-dependent translation of $u$, which yields a non--isotropic oscillator metric
\[
G=2dUdT-K^2 Z^2 dT^2-dZ^2-dX^2-dY^2.
\]
This metric is not conformally flat, but it
has a constant curvature. At the level of the Schwarzchild--dS
space the cosmological horizon has been shifted. This, as discussed in \cite{GHds} (see also \cite{witten}), leads to an ambiguity in a definition of a Hilbert
space which is observer dependent. In our work this manifests
itself in the ambiguity of quantum phase.
\subsection{The Penrose effect from the Unruh effect}
\label{secu}
Let us  go back to the uniform gravitational field with $\Lambda=0$. We will show that the coefficient of $t^3$ in the phase ambiguity (\ref{phase1}) is related
to the surface gravity in the Rindler horizon of an accelerating observer, and argue that the phase-shift is a
non--relativistic counterpart of non--uniqueness of vacuum in the Unruh effect \cite{unruh}.

The Kottler-M\"oller
transformation 
\begin{eqnarray}
\label{KMt}
T&=&-\frac{1}{c}\Big(z-\frac{c^2}{\gamma}\Big)\sinh\Big(\frac{\gamma t}{c}\Big),\\
X&=&x,\quad Y=y, \quad  Z=\Big(z-\frac{c^2}{\gamma}\Big)
\cosh\Big(\frac{\gamma t}{c}\Big)+\frac{c^2}{\gamma},\quad\mbox{where}\quad \gamma=\mbox{constant}\nonumber
\end{eqnarray}
applied to  the Minkowski metric $c^2 dT^2-dX^2-dY^2-dZ^2$
pulls it  back to
\be
\label{gmetric}
  g=\frac{(c^2-\gamma z)^2}{c^2} dt^2-dx^2-dy^2-dz^2
\ee
with the Killing horizon at $z=c^2/\gamma$, and the surface gravity $\kappa=\gamma/c$.
The quantisation of the scalar field $F$ satisfying the massive Klein--Gordon equation
\be
\label{KG}
  \Box F=\frac{m^2c^2}{\hbar^2} F
\ee
on  the $g$--background involves using the Killing vector $\p_t$ of the accelerating observer to define the positive and negative frequency modes. These modes will differ from those obtained on the Minkowski background w.r.t the Killing vector $\p_T$ of the inertial observer.
Computing the Bogoliubov transformation between the two set of modes shows
that the particle number is observer dependent. A vacuum state of the inertial observer corresponds to particle creation w.r.t the accelerating observer
moving along the world-lines  generated by $\p_t$. This is the Unruh effect \cite{unruh}.

Now consider the non--relativistic limit of (\ref{gmetric}), and apply the limiting procedure to solutions of the 
Klein--Gordon equation (\ref{KG}). The metric blows up in the limit $c\rightarrow \infty$ but its Christoffel symbols stay finite
\[
  \Gamma^t_{tz}=\Gamma^t_{zt} =\frac{-\gamma}{c^2-\gamma z}, \quad
  \Gamma^z_{tt}=-\frac{(c^2-\gamma z)\gamma}{c^2}.
\]
The Newtonian connection $
\Gamma^z_{tt}=-\gamma
$  resulting from the $c\rightarrow\infty$ limit 
is the gradient of the potential corresponding to the uniform gravitational field. 
Making the ansatz $
  F=e^{-imc^2t/\hbar} \psi(z, t)
$ in (\ref{KG}) gives, in $c\rightarrow \infty$ limit,  the Schr\"odinger equation
\[
  i\hbar\frac{\p \psi}{\p t}=-\frac{\hbar^2}{2m}\frac{\p^2 \psi}{\p z^2}-m\gamma z \psi.
\]
The wave function $\psi$ is unitary equivalent to that of the free Schr\"odinger equation, but with a non--linear phase shift 
(\ref{phase1}). This, as we have already discussed, also leads to the ambiguities in the negative/positive frequency decomposition. Moreover the $c\rightarrow \infty$ limit of the  Kottler-M\"oller transformation (\ref{KMt})
is  (\ref{coords1}).
Thus the Penrose effect described in \cite{Pe11} is a non--relativistic limit of the Unruh effect \cite{unruh}. 
\section{Twistor realisation of the Newton--Hooke group.}
\label{section2}
The de--Sitter  space is conformally flat, and its twistor space coincides with that
of the Minkowski space \cite{PR, todds}. The difference lies in fixing the conformal scale, and at the twistor level can be seen by considering the Klein correspondence:
A line in the projective twistor space $\CP^3$ with homogeneous coordinates $[Z^{\alpha}], \alpha =1, \dots, 4$ is represented by two points (twistors) $X$ and $Y$ on this line, or by a bi--vector defied up to an overall scale
$
P^{\alpha\beta}=X^{\alpha}Y^{\beta}-X^{\beta}Y^{\alpha}
$
which is  simple, i. e.
\be
\label{quadric}
\epsilon_{\alpha\beta\gamma\delta}P^{\alpha\beta}P^{\gamma\delta}=0.
\ee
The space of lines is the Klein quadric  ${\mathcal Q}$, given by (\ref{quadric}),  in the projective space 
$\CP^5$.
This gives an identification between ${\mathcal Q}$ 
and the complexified, compactified Minkowski space: Lines in $\CP^3$ correspond to points in ${\mathcal Q}$, and points in $ \CP^3$ correspond to holomorphic 2-planes
in ${\mathcal Q}$ (called the $\alpha$--planes) given by the locus of points $P^{\alpha\beta}$ such that
$
\epsilon_{\alpha\beta\gamma\delta}Z^{\alpha}P^{\beta\delta}=0.
$
The conformal class on ${\mathcal Q}$ is defined by declaring
two simple bi--vectors $P$ and $Q$ to be null separated, if the corresponding lines intersect at a point in the twistor space.
For any fixed $P$ this condition defines a tangent plane  to ${\mathcal Q}$
which intersects $\mathcal{Q}$ in a cone. This is the quadratic condition for the light--cone
of $P$ in the complexifed and compactified Minkowski space. 

To select  a metric of  constant 
curvature in this conformal class pick a  bi--vector  $I$, from now on called
the {infinity twistor}, which does not belong to ${\mathcal Q}$. Then, for any two points
$P, Q$ in ${\mathcal Q}$, there exists the unique plane in $\CP^5$ containing $I$, and these two points. This plane
intersects the polar hyper--plane of $I$ in projective line, which in turn meets the Klein quadric ${\mathcal Q}$ in two points, $A$ and $B$. The distance between $P$ and 
$Q$ is then defined by the cross ratio of four points \cite{HT}.
\[
d(P, Q)=\frac{1}{2}\mbox{ln}|P, Q, A, B|.
\]

We represent the infinity twistor $I$ as a contact one--form
\be
\label{contact_form}
\theta=\frac{1}{2}\epsilon_{\alpha\beta\gamma\delta}I^{\gamma\delta}Z^{\alpha}dZ^{\beta}=\pi_{A'}d\pi^{A'}-\Lambda \omega_Ad\omega^A
\ee
where, in 
Penrose's two-component spinor notation \cite{PR}, $[Z]=[\omega^0,\omega^1, 
\pi_{0'}, \pi_{1'}]$.
The action of the  conformal group $SL(4, \C)$ on the complexified and compactified Minkowski space 
$M_{\C}={\mathcal Q}$
maps
$\alpha$--planes to $\alpha$--planes. Thus it extends to a holomorphic projective action on the 
twistor space $\CP^3$ generated by the linear
vector fields
take the form
\be
  \label{KdS}
  {\mathcal K}= \phi^A_B\omega^B\frac{\p}{\p \omega^A}+
  {{{{\tilde\phi}}^{A'}}_{B'}}\pi_{A'}\frac{\p}{\p \pi_{B'}}+
                  P^{AA'}\pi_{A'}\frac{\p}{\p \omega^A}+B_{A'B}\omega^B\frac{\p}{\p \pi_{A'}}
  +C\Big(\pi_{A'}\frac{\p}{\p \pi_{A'}}-\omega^A\frac{\p}{\p \omega^A}\Big).    
\ee
Here $P^{AA'}$ generates  translations, the symmetric spinors
$\phi_{AB}=\phi_{(AB)}$ and $\phi_{A'B'}=\phi_{(A'B')}$ are ASD and SD parts of the rotation
$M_{ab}$, the scalar $C$ is the  dilatation, and $B^{AA'}$ generate special conformal transformations.
The subgroup of $SL(4, \C)$ preserving the non--simple de--Sitter infinity twistor $I=d\theta$
is the complexified de-Sitter group $SO(5, \C)$. The condition ${\mathcal L}_K\theta=0$ gives
$B_{AA'}=\Lambda P_{AA'}$, and $C=0$
which reduces the 15 parameters down to 10.
In what follows, we shall perform the In\"onu--Wigner contraction of the twistor de--Sitter Lie algebra to the the 
Newton--Hooke algebra \cite{LB} where $c\rightarrow \infty, \Lambda\rightarrow 0$, and
$\omega^2=\frac{1}{3}c^2\Lambda$ is fixed.  
\vskip5pt
We will first recall the construction of a non-relativistic twistor space for the flat Newtonian 
space time \cite{DG16, ADM17}, and then  introduce the non--relativistic infinity twistor. The starting point is to consider family of relativistic projective twistor spaces $PT_c= \CP^3\setminus \CP^1$ parametrised by the speed of light $c$.

Let $T^{AA'}=\frac{1}{\sqrt{2}}(o^{A}o^{A'}+\iota^{A}\iota^{A'})$ be a unit vector in $M_{\C}$ 
defining a $3+1$ split, and let $U$ and $\widetilde{U}$ be open sets on $PT_c$ corresponding to
$\pi_{1'}\neq 0$ and $\pi_{0'}\neq 0$ respectivelly. We use $T^{AA'}$ to 
 define inhomogeneous coordinates $(Q, T)$ and $(\widetilde{Q}, \widetilde{T})$
on $U$ and $\widetilde{U}$ in $PT_c$ by
\be
\label{patch_U}
\Big(Q=2\frac{ {T_A}^{A'}\omega^A\pi_{A'} }{{\pi_{1'}}^2}, 
T=  \frac{\sqrt{2}}{c}\frac{\omega^1}{\pi_{1'}}, \lambda=\frac{\pi_{0'}}{\pi_{1'}} \Big), \quad
\Big(\widetilde{Q}=2\frac{ {T_A}^{A'}\omega^A\pi_{A'} }{{\pi_{0'}}^2}, 
\widetilde{T}=
\frac{\sqrt{2}}{c}
\frac{\omega^0}{\pi_{0'}}, \tilde{\lambda}=\frac{\pi_{1'}}{\pi_{0'}} \Big) 
\ee
with the patching relation on $U\cap\widetilde{U}$
\[
\tilde{\lambda}=\frac{1}{\lambda}, \quad
\left (
\begin{array}{cc}
\widetilde{T}\\
\widetilde{Q} 
\end{array}
\right )=F_c
\left (
\begin{array}{cc}
{T}\\
{Q} 
\end{array}
\right ) \quad\mbox{where}\quad F_c=\left (
\begin{array}{cc}
1 & -{(c\lambda)}^{-1} \\
0 & {\lambda^{-2}}
\end{array}
\right ),
\]
and the twistor/space--time incidence relation
\[
Q= -(x+iy)-2\lambda z+\lambda^2(x-iy), 
\quad T=t-\frac{1}{c}(z-\lambda(x-iy)).
\]
For finite, and non--zero $c$
there exist $H(\lambda), \widetilde{H}(\tilde{\lambda}) \in GL(2, \C)$ such that
$F_c=\widetilde{H}\;\mbox{diag}(\lambda^{-1}, \lambda^{-1})\; H^{-1}$. For $c=\infty$ we have $F_{\infty}=\mbox{diag}(1 , \lambda^{-2})$. Therefore, in the limit, the holomorphic type of the twistor lines jumps, and
\[
PT_c=\OO(1)\oplus\OO(1) \quad\mbox{for}\quad c<\infty\quad\mbox{and}\quad PT_\infty=\OO\oplus\OO(2).
\]
Expressing the infinity twistor (\ref{contact_form})
in the twistor coordinates $(Q, T, \pi_{A'})$
gives
\[
\theta= \Big(1-\frac{1}{2}\Lambda c^2 T^2\Big)\pi_{A'}d\pi^{A'}-\frac{\Lambda c}{2} 
(\pi_{1'})^2 (TdQ-QdT)
\]
with  the Newton--Hooke limit
\[
\theta_0=
\Big(1-\frac{3\omega^2}{2}T^2\Big)\pi_{A'}d\pi^{A'}.
\]
We now push forward the vector fields (\ref{KdS}) with $B=\Lambda P, C=0$
by the transformation (\ref{patch_U}). Before implementing the limiting procedure 
the Hamiltonian $H$, which is the  $T_{AA'}$ component of ${\mathcal K}$ should be multiplied by $c$, and the $T_{AA'}$ component
of $M_{ab}$ should be divided by $c$. The In\"onu--Wigner contraction then yields a 10--dimensional algebra of global vector fields. We represent these in the coordinate
patch with $\lambda=\pi_{0'}/\pi_{1'}\neq \infty$ as
\begin{eqnarray*}
  H&=&\p_T-\frac{3\omega^2}{2}T(T\p_T+2Q\p_Q), \\
  P_1&=&-i\Big(1+\frac{3\omega^2}{2}T^2\Big)\lambda\p_Q, \;
  P_2= -\frac{1}{2}
         \Big(1+\frac{3\omega^2}{2}T^2\Big)
         (1+\lambda^2)\p_Q, \;
  P_3=-\frac{i}{2}\Big(1+\frac{3\omega^2}{2}T^2\Big)
         (1-\lambda^2)\p_Q\\
  K_1&=&-iT\lambda\p_Q,\;
  K_2=-\frac{1}{2}T(1+\lambda^2)\p_Q,\;
  K_3=-\frac{i}{2}T (1-\lambda^2)\p_Q\\
  J_1&=& -i(\lambda\p_\lambda+Q\p_Q), \;
  J_2=-\frac{1}{2}(1+\lambda^2)\p_\lambda-\lambda Q\p_Q,\; 
  J_3=-\frac{i}{2}(1-\lambda^2)\p_\lambda+i\lambda Q\p_Q. 
\end{eqnarray*}
These vector fields form a subalgebra of the infinite--dimensional algebra $\mathfrak{h}$ of global holomorphic vector fields on $PT_\infty$ considered in \cite{gundry} (see \cite{kosinski} for different subalgebra)
\[
[H, P_i] = 6\omega^2 K_i, \quad [H, K_i]=P_i,\quad [J_i, J_j] = \epsilon_{ijk} J_k, \quad
[J_i, P_j]=\epsilon_{ijk} P_k, \quad
[J_i, K_j]=\epsilon_{ijk} K_k
\]
This is the Newton--Hooke algebra \cite{LB}. Its ten generators are the rotations $J_i$, the translations $P_i$, the boosts $K_i$ and the Hamiltonian $H$. Its central extension
is the algebra of isometries of the Eisenhart metric (\ref{cord_change}) of the non--relativistic de-Sitter space.

\subsubsection*{Acknowledgements}
I am grateful to Gary Gibbons, Peter Horvathy and Roger Penrose for their insightful comments.

\end{document}